# Edgar Allan Poe: the first man to conceive a Newtonian evolving Universe

Paolo Molaro
INAF - Osservatorio Astronomico di Trieste, Italy
Alberto Cappi
INAF - Osservatorio Astronomico di Bologna, Italy
Université de Nice Sophia-Antipolis, CNRS, Observatoire de la Côte d'Azur,
UMR 6202 Cassiopée, BP 4229, F-06304 Nice Cedex 4, France

**Abstract.** The notion that we live in an evolving universe was established only in the twentieth century with the 'discovery' of the recession of galaxies by Hubble and with the Lemaitre and Friedmann's interpretation in the 20s. However, the concept of an evolving universe is intrinsically tied to the law of universal gravitation, and it is surprising that it remained unrecognized for more than two centuries. A remarkable exception to this lack of awareness is represented by Poe. In *Eureka* (1848), the writer developed a conception of an evolving universe following the reasoning that a physical universe cannot be static and nothing can stop stars or galaxies from collapsing on each other. Unfortunately this literary work was, and still is, very little understood both by the literary critics and scientists of the time. We will discuss Poe's cosmological views in their historical scientific context, highlighting the remarkable insights of the writer, such as those dealing with the Olbers paradox, the existence of other galaxies and of a multi-universe.

> *Deep into the darkness peering, long I stood*
> *there, wondering, fearing,*
> *Doubting, dreaming dreams no mortal ever*
> *dared to dream before.[1]*

The concept of an evolving Universe originated from the discovery of the recession of galaxies by Hubble, and with the Lemaitre and Friedmann solutions of Einstein's equations in the late 1920s. Later in the 1960s it found a definitive acceptance as the Big Bang theory with the detection of cosmic background radiation and theoretical-observational concordance of primordial nucleosynthesis. However, the necessity of an evolutionary physical universe is intrinsically tied to the Newtonian law of gravitation, a fact that, quite surprisingly, remained unrecognized for more than two centuries. A remarkable exception to this unawareness is represented in the book *Eureka* by Edgar Allan Poe in 1848, where the idea of a collapsing universe is elaborated for the first time in Newtonian physics. A physical universe cannot be static and Poe accepted the idea that nothing is stopping stars from collapsing onto each other. Starting from metaphysical assumptions, through logical reasoning he developed what is the first concept of an evolving universe within Newtonian physics. This book was very little understood both by the literary critics and by scientists of his times, and to some extent this is still true today. Here we will describe Poe's cosmological views in their historical context, highlighting their remarkable insights such as those dealing with a finiteness of the universe, the nature of the nebulae, the shape of the Milky Way, the existence of other galaxies and that of a multi-universe. Eureka is a rare example of a literature that preceded and advanced the science of its times and can be regarded as the first book of Newtonian cosmology.

---

[1] From the poem The Raven, E.A. Poe 1845.

**Introduction**

Milne and McCrea showed in 1934 that a Newtonian Cosmology is in many respects equivalent to relativistic cosmology. The equation for a particle with mass moving in a finite cloud with uniform density is formally similar to that obtained by using general relativity (see Tipler 1996 for a recent account). In particular the expansion of the Universe, and therefore the necessity of a Big Bang, is not a specific feature of General Relativity but follows directly from the application of the Newtonian law formulated back in 1687. What happened in between? Why was Newtonian cosmology only derived formally after relativistic cosmology?

A remarkable exception to this unawareness is represented by Edgar Allan Poe, who, in 1848 in his last poem Eureka arrived at a sort of Newtonian Cosmology.

Edgar Allan Poe (1809-1849), renowned writer, poet, editor and literary critic, inventor of the detective-fiction genre, and best known for his tales of mystery and macabre left an enormous legacy in several fields. Here we argue that he also made important steps in science that unfortunately have gone unrecognized. In the following we sketch the story of this remarkable book. For further reading refer to Cappi (1994).

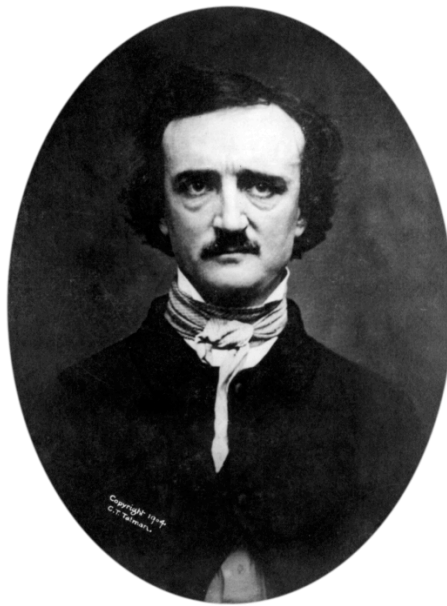 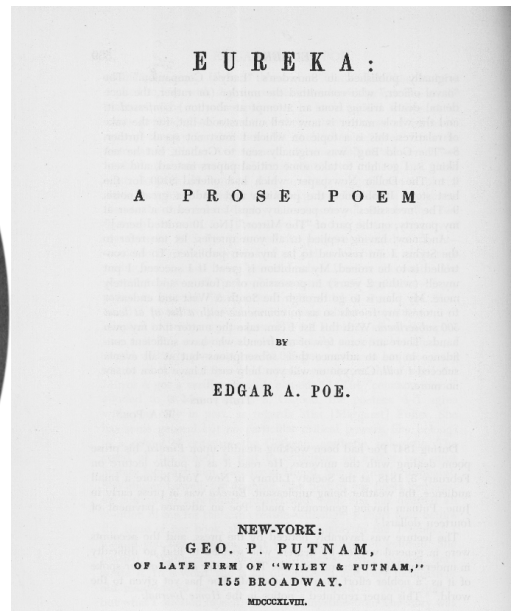

Fig1. Daguerreotype of Poe and 1848 cover of Eureka

**The lecture**

In the year 1848 Poe planned a promotional tour of the Southern and Western states, the purpose of which was to raise the necessary funds to launch the magazine Stylus by delivering public lectures. For one of these he organized his scientific and astronomical views.

Mrs. Clemm recalls that preparatory phase: *"He [Poe] never liked to be alone, and I used to sit up with him, often until four o'clock in the morning, he at his desk, writing, and I dozing in my chair.*

*When he was composing Eureka, we used to walk up and down the garden, his arm around me, mine around him, until I was so tired I could not walk. He would stop every few minutes and explain his ideas to me, and ask if I understood him.* [2] The first lecture happened on the 2-3 February 1848. The Daily Tribune of New York carried the advertisement: *"Edgar A. Poe will lecture at the Society Library on Thursday evening, the 3d inst at half-past 7. Subject, 'The Universe.' Tickets 50 cents — to be had at the door.[3]"*

We can have an idea of that lecture through the words of Maunsell B. Field, a young lawyer, that was in the audience and recalls the conference: *It was a stormy night, and there were not more than sixty persons present in the lecture-room. I have seen no portrait of POE that does justice to his pale, delicate, intellectual face and magnificent eyes. His lecture was a rhapsody of the most intense brilliancy. He appeared inspired, and his inspiration affected the scant audience almost painfully. He wore his coat tightly buttoned across his slender chest; his eyes seemed to glow like those of his own raven, and he kept us entranced for two hours and a half* [4]. But is also clear from the days after reports in several newspapers that nobody understood what Poe was talking about. Feeling misunderstood, Poe expanded the lecture into a book. Around 11th of July the New York publisher George Putnam issued 500 copies of Eureka. Poe was really convinced to have found something very important as it can be inferred from the account of the encounter between the writer and the publisher Putnam in the latter's office at 155 Broadway:

*Seated at my desk, and looking at me a full minute with his glittering eye, he at length said: I am Mr. Poe. I was all ear; of course, and sincerely interested. It was the author of The Raven, and of The Gold Bug! I hardly know, said the poet, after a pause, how to begin what I have to say. It is a matter of profound importance. After another pause, the poet seeming to be in a tremor of excitement, he at length went on to say that the publication he had to propose was of momentous interest. Newton's discovery of gravitation was a mere incident compared to the discoveries revealed in this book. It would at once command such universal and intense attention that the publisher might give up all other enterprises, and make this one book the business of his lifetime. An edition of fifty thousand copies might be sufficient to begin with.[5]*

**A finite Universe**

Astronomy in the mid 1800s was essentially observational. It was the astronomy of the John Herschel and Friedrich Bessel dealing mainly with the study of comets, binary stars, nebulae, star counts and stellar distances. In 1838 Bessel was the first to measure stellar distances, the planet Neptune was discovered in 1846 by Galle on the predictions of Le Verrier, and the 72-inch Newtonian reflector of Lord Rosse in 1845 resolved the Orion nebula and a few years later imaged the spiral galaxy M101. Poe quite differently considers the Universe as a whole, trying to understand its origin and fate.

*I do not know a treatise in which a survey of the Universe -- using the word in its most comprehensive and only legitimate acceptation -- is taken at all.*[6] According to Poe himself among the

---

[2] R.E. Shapley in a Philadelphia newspaper, quoted by George E. Woodberry (1894).
[3] Dwight R. Thomas and David K. Jackson *The Poe Log* (1987), Chapter 10, p 713
[4] Maunsell B. Field Memories of Many Men and Some Women 1873, pp. 224-225
[5] George Palmer Putnam, "Leaves from a Publisher's Letter-Book," *Putnam's Monthly*, October 1869, pp. 471.
[6] E u r e k a :  a  P r o s e  P o e m **.** by Edgar A. Poe**.** New-York: Geo. P. Putnam, of late Firm of "Wiley & Putnam," 155 Broadway. MDCCCXLVIII. p. 9

contemporaneous books the one which is more similar is the Cosmos by Alexander Von Humboldt, to whom the poem is dedicated *with very profound respect.*

Poe's approach is the one of modern cosmology which make a global model out of few relevant observational facts. Not without a certain amount of snobbish attitude Poe remarks: *The error of our progenitors was quite analogous with that of the wiseacre who fancies he must necessarily see an object the more distinctly, the more closely he holds it to his eyes. They blinded themselves, too, with the impalpable, titillating Scotch snuff of detail.*[7]

Poe's crucial starting point was to abandon the assumption of an infinite universe. For Poe *the whole idea of unlimited Matter is not only untenable, but impossible and preposterous*[8].

First on somewhat philosophical grounds: *as an individual, I may be permitted to say that I cannot conceive Infinity, and am convinced that no human being can. A mind not thoroughly self-conscious -- not accustomed to the introspective analysis of its own operations – will*; secondly on empirical grounds: *all observation of the firmament refutes -- the conception of the absolute infinity of the Universe of stars.*[9]

Poe makes use of the Kepler's argument against the infinity of the universe, nowadays known as Olbers paradox. Kepler, in his reply to the publication of the "Nuncius Sidereus" of March 1609, where Galileo reported the discovery of new stars, wrote: *You [Galileo] do not hesitate to declare that there are visible over 10000 stars. The more they are, and more crowded they are, the stronger becomes my argument against the infinity of the universe, as set forth in my book "The New Star"*[10]. Poe does not refer explicitly to Kepler but makes the same reasoning. Departing from Olbers's solution, which was adopted by Von Humboldt and other contemporaneous scientists, he shows considerable autonomy of thinking and self-confidence. Poe writes:

*observation assures us that there is, in numerous directions around us, certainly, if not in all, a positive limit [...] Were the succession of stars endless, then the background of the sky would present us an uniform luminosity, like that displayed by the Galaxy -- since there could be absolutely no point, in all that background, at which would not exist a star.*[11]

Thus a distinction between space and the stellar system is introduced and Poe's concept of the universe is much similar to the universe of the Stoics as elaborated by Zeno di Citium (334-262 BC), albeit with an important difference which we will see shortly:

*the Universe of stars has always been considered as coincident with the Universe proper, [ ...] It has been always either directly or indirectly assumed -- at least since the dawn of intelligible Astronomy -- that, were it possible for us to attain any given point in space, we should still find, on all sides of us, an interminable succession of stars. This was the untenable idea of Pascal when making perhaps the most successful attempt ever made, at periphrasing the conception for which we struggle in the word "Universe." "It is a sphere," he says, "of which the centre is everywhere, the circumference, nowhere.*[12]

From a finite universe Poe conceived that the number of stars was less than the atoms composing a cannon ball: *in a wilderness of atoms so numerous that those which go to the composition of a cannon-ball, exceed, probably, in mere point of number, all the stars which go to the constitution of the Universe*[13] - a statement that we know to be true today but that certainly was quite difficult to conceive in the mid-XIX century. However, Poe's inspiration was to apply the gravitational force into a finite universe.

---

[7] Ibid., p. 13
[8] Ibid., p. 69
[9] Ibid., p. 26
[10] J. Kepler *Dissertatio cum Nuncio Sidereo* (*Conversation with the Starry Messenger*) 1610
[11] E u e k a by E. A. Poe. op. cit. p. 100
[12] Ibid., p 28. This definition of the universe was not invented by Pascal: it had already been used by Giordano Bruno, and in the Middle Ages it was used to define not the universe but God.

**An evolving Universe**

Gravity for Poe is the central force modeling the physical world. In the book he quotes the Maskelyne, Cavendish and Bailly experiments that measured the gravitational attraction of the mass of a mountain and found it to be consistent with the Newton's law. Poe goes further in trying to grasp the deep nature of gravitation:

*What does the Newtonian law declare? –[...] Every atom, of every body, attracts every other atom, both of its own and of every other body, with a force which varies inversely as the squares of the distances between the attracting and attracted atom. – Here, indeed, a flood of suggestion bursts upon the mind. [...] There is nothing to impede the aggregation of various unique masses, at various points of space*[14].

These considerations recall those of Bentley's correspondence of 1692-1693. The theologian Richard Bentley wanted make a confutation of atheism by using the Newtonian law and wrote to Newton several letters asking what would be the effect of gravity on matter evenly scattered throughout an infinite space. Newton's replies were that stars were spread throughout infinite space with near uniformity, so that each star is balanced between equal and opposite gravitational pulls. However, he admitted that only if matter were distributed with perfect and unbroken uniformity would every particle remain at rest. Subsequently on the 5 of May 1694 in a private letter to Gregory, Newton admitted that: *A continual miracle is needed to prevent the Sun and the fixed stars from rushing together through gravity*. God is seen not only as a passive clockmaker but is continuously acting on the stellar universe to prevent it from collapsing. Note that William Stukeley brought to the attention of Newton and Halley the riddle of the darkness of night-sky in connection with the problem of infinity (Hoskin 1985).
Poe, free from theological prejudices, just left the matter evolving naturally according to the Newtonian law:

*the assertion that "if there were but one body in the Universe, it would be impossible to understand how the principle, Gravity, could obtain": -- [...] That so pregnant a suggestion as the one quoted should have been permitted to remain so long unfruitful, is, nevertheless, a mystery which I find it difficult to fathom.*[15]

We do not know the source of the statement he quoted, but here Poe found the key to his vision; gravity is the force which acts to bring matter back to the original unity: *The tendency of the diffused atoms to return into their original unity, would be understood as the principle of the Newtonian law of gravity*

---

[13] Ibid., p. 42
[14] Ibid, p. 39
[15] Ibid., p. 129

*[...] Gravity exists on account of Matter's having been irradiated.*[16]
Thus matter was initially irradiated by a sort of *flash*, followed by an expansion, and subsequently a contraction under the action of gravity.

**Dimensions, Space and Voids**

Poe takes the size the Universe of about 3 million light years, as estimated by William Herschel from the limiting magnitude of the objects he was able to see with his telescope: *yet so far removed from us are some of the "nebulae" that even light, speeding with this velocity, could not and does not reach us, from those mysterious regions, in less than 3 millions of years. This calculation is made by the elder Herschel.*[17] This was an overwhelmingly large size for those days, which lacked consensus, and was abandoned by Herschel himself in later years. In Poe's view the size of the universe is the result of the cosmological irradiation and the large size of the universe is required to allow the formation of stars, solar systems and life. This accounts also for the existence of the observed extremely large voids:

*the difficulty which we have so often experienced, while pursuing the beaten path of astronomical reflection, in accounting for the immeasurable voids alluded to – in comprehending why chasms so totally unoccupied and therefore apparently so needless, have been made to intervene between star and star – between cluster and cluster – in understanding a sufficient reason for the Titanic scale, in respect of mere Space, on which the Universe is seen to be constructed. A rational cause for the phaenomenon, I maintain that Astronomy has palpably failed to assign: – but the considerations through which, in this Essay, we have proceeded step by step, enable us clearly and immediately to perceive that Space and Duration are one*[18].

Poe clearly understands the distance implied by a light year and the implications of light travel time:

*There are "nebulae" which, through the magical tube of Lord Rosse, are this instant whispering in our ears the secrets of a million of ages by-gone. In a word, the events which we behold now -- at this moment -- in those worlds -- are the identical events which interested their inhabitants ten hundred thousand centuries ago. ...this suggestion forces upon the soul -- rather than upon the mind --*[19]

**The nature of nebulae**

In the XIX century the nature of nebulae was actively debated. Are Nebulae nearby gaseous objects or distant stellar systems? When Lord Rosse in 1845 resolved stars in Orion many astronomers abandoned the interpretation that stars form from nebulae. John Herschel in Outlines of Astronomy (1849) wrote that nebulae are cluster of stars in the Galaxy and Agnes Clerk in "System of Stars" of 1890 wrote that the question whether the nebulae are external galaxies hardly any longer needs discussion they being all galactic. The debate ended when the extragalactic nature of some nebulae was established in 1924 by Edwin Hubble.

---

[16] Ibid., p. 38
[17] Ibid., p. 117
[18] Ibid., p. 117
[19] Ibid., p. 117

At variance with the opinion of many astronomers of his times, Poe supported the extragalactic nature of the nebulae and found an explanation for the new observations of Lord Rosse in Orion - the nebulae did not contain gas just because they had formed stars in the past. It is worth following closely this passage:

*A most unfounded opinion has been latterly current and even in scientific circles -- the opinion that the so-called Nebular Cosmogony has been overthrown. This fancy has arisen from the report of late observations made, among what hitherto have been termed the "nebulae," through the large telescope of Cincinnati, and the world-renowned instrument of Lord Rosse. Certain spots in the firmament which presented, even to the most powerful of the old telescopes, the appearance of nebulosity, or haze, had been regarded for a long time as confirming the theory of Laplace. They were looked upon as stars in that very process of condensation which I have been attempting to describe. Thus it was supposed that we "had ocular evidence" -- an evidence, by the way, which has always been found very questionable -- of the truth of the hypothesis; the most interesting was the great "nebulae" in the constellation Orion: -- but this, with innumerable other miscalled "nebulae," when viewed through the magnificent modern telescopes, has become resolved into a simple collection of stars. Now this fact has been very generally understood as conclusive against the Nebular Hypothesis of Laplace [...]*[20]
*Through what we know of the propagation of light, we have direct proof that the more remote of the stars have existed, under the forms in which we now see them, for an inconceivable number of years. So far back at least, then, as the period when these stars underwent condensation, must have been the epoch at which the mass-constitutive processes began. That we may conceive these processes, then, as still going on in the case of certain "nebulae," while in all other cases we find them thoroughly at an end.*[21]

These nebulae, which are other galaxies as our own, are seen to be in collapse: *A nucleus is always apparent, in the direction of which the stars seem to be precipitating themselves; nor can these nuclei be mistaken for merely perspective phaenomena: -- the clusters are really denser near the centre -- sparser in the regions more remote from it. In a word, we see every thing as we should see it were a collapse taking place.* [22]He also ventures to argue against John Herschel's ideas: To *"Herschel [John] there is evidently a reluctance to regard the nebulae as in "a state of progressive collapse."* [23]
*"why is he disinclined to admit it? Simply on account of a prejudice; -- merely because the supposition is at war with a preconceived and utterly baseless notion -- that of the endlessness -- that of the eternal stability of the Universe.*[24]

The view of collapsing clusters of stars, which we now know to be imprecise, reflects the new way Poe is looking at the universe, as an evolving entity together with its constituents.

**Other Poe's insights**

We list here briefly some of the poet's views which reveal a fine mind and considerable prescience. The shape of the Milky Way was believed highly asymmetric with two arms forming a big Y. Poe instead

---

[20] Ibid., p. 88
[21] Ibid., p. 90
[22] Ibid., p. 128
[23] Ibid., p. 25
[24] Ibid., p. 128

favored a rather symmetric shape starting from a solar system analogy:

*There has been a great deal of misconception in respect to the shape of the Galaxy; which, in nearly all our astronomical treatises, is said to resemble that of a capital Y. The cluster in question has, in reality, a certain general -- very general resemblance to the planet Saturn with its encompassing triple ring[25].*

Poe envisaged a correct structure of the large scale of the universe. The universe is composed of a myriad of systems, as is our own Galaxy, assembled in condensations by gravity, leaving large voids between:

*The Universe exists as a roughly spherical cluster of clusters, irregularly disposed[26]. The smaller systems, in the vicinity of a larger one, would, inevitably, be drawn into still closer vicinity. A thousand would assemble here; a million there -- perhaps here, again, even a billion -- leaving, thus, immeasurable vacancies in space.[27]*

Poe makes a connection between the formation of the solar system according to the Laplace theory and the evolution of life on Earth. Life is participating in the universal evolution. In fact Poe models the universe to allow the growth of life and human evolution:

*We thus reach the proposition that the importance of the development of the terrestrial vitality proceeds equably with the terrestrial condensation. [...] Now this is in precise accordance with what we know of the succession of animals on the Earth. As it has proceeded in its condensation, superior and still superior races have appeared.[28]"*

Poe admitted the existence of giant stars, dark stars and in general the existence of a larger variety of celestial objects: *"There is the very best reason for believing that many of the stars are even far larger than the one we have imagined*, i.e. much bigger than the solar system. [29]*" we know that there exist non-luminous suns -- that is to say, suns whose existence we determine through the movements of others, but whose luminosity is not sufficient to impress us[30]*.
In a passage he seems to anticipate Mach's Principle (1883) which states that the inertia of a body is the result of the interaction with all the other bodies of the universe.

*If I propose to ascertain the influence of one mote in a sunbeam upon its neighboring mote, I cannot accomplish my purpose without first counting and weighing all the atoms in the Universe and defining the precise positions of all at one particular moment. If I venture to displace, by even the billionth part of an inch, the microscopical speck of dust which lies now upon the point of my finger, what is the character of that act upon which I have adventured? I have done a deed which shakes the Moon in her path, which causes the Sun to be no longer the Sun, and which alters forever the destiny of the multitudinous myriads of stars that roll and glow in the majestic presence of their Creator.* [31]

---

[25] Ibid., p. 97
[26] Ibid., p. 96
[27] Ibid., p. 95
[28] Ibid., p. 85
[29] Ibid., p. 111
[30] Ibid., p. 84
[31] Ibid., p. 42

4The analogy with Mach's Principle has been also noted by Alpher and Herman in their book "Genesis of the Big Bang" (2001).

In a passage at the end of the book Poe wrote about a multi-verse (similar to the inflationary multi-verses postulated today):

*As an individual, I myself feel impelled to the fancy [...] that there exist a limitless succession of Universes more or less similar to that we have cognizance.* Add to this: *If such clusters of clusters exist, however -- and they do -- it is abundantly clear that, having had no part in our origin, they have no portion in our laws. They neither attract us, nor we them. [...] Among them and us -- considering all, for the moment, collectively -- there are no influences in common.*[32]

**Conclusions**

Eureka is a unique book. It does not only incorporate astronomy, but is an astronomical book written by a poet with deep scientific insights. Putting aside the metaphysical introduction and the lyric conclusion, in its central part Eureka is nearly a textbook of theoretical Newtonian cosmology, but without mathematics. As a matter of fact, E.A. Poe has been the first man to imagine an evolving Universe in a Newtonian frame, which is not much different from modern views, thus introducing a sort of inverse Big Bang model: *what God originally [...] created nothing but Matter in its utmost conceivable state of Simplicity. [...] the primordial Particle.*[33]

We do not know of any scientist who openly found inspiration from Poe's book. However, George Lemaître (1931), who did the first step towards the Big Bang theory, wrote something very similar: *If we go back in the course of time we must find fewer and fewer quanta, until we find all the energy of the universe packed in a few or even in a unique quantum*, which he called the Primeval Atom*,* and Poe was well known in the French spoken world since Paul Valéry wrote a pamphlet *Au sujet of Eureka* in 1924 and Baudelaire had translated *Eureka* into French as early as 1864.

Moreover, in a biography of Alexander Friedmann, the scientist that probably first arrived at the idea of an evolving universe (Tropp et al. 1993), it is reported that Friedmann's preferred authors were Dostoevsky, Hoffman and... Poe. But we have to stop here and leave these speculations to the reader.

**Acknowledgements**
We warmly thank Charles Marussich for editing the English text.**References**

Alpher, R.A. and Herman, R., *Genesis of the Big Bang* (2001, Oxford University Press).

Cappi, A., 'Edgar Allan Poe's Physical Cosmology', Quarterly Journal of the Royal Astronomical Society, Vol.35 (1994), pp.177-192.

Harrison, E.R., *Darkness at Night: A Riddle of the Universe* (1987, Harvard University Press).---

[32] Ibid., p. 03
[33] Ibid., p. 29

Hoskin, M., 'Stukeley's Cosmology and the Newtonian Origins of Olber's Paradox', Journal for the History of Astronomy, 16 (1985), p.77.

Lemaitre, G., 'The Beginning of the World from the Point of View of Quantum Theory', Nature Vol 127 (1931), p.706.

Milne, W.H., 'A Newtonian Expanding Universe', Quarterly Journal Math. Oxford, Vol.5 (1934), pp.64-72.

Milne, W.H. and McCrea E.A., 'Newtonian Universes and the curvature of space', Quarterly Journal Math. Oxford, Vol. 5 (1934), pp.73-80.

Tropp, E.A., Frenkel, V. Ya, and Chernin, A.D., *Alexander A. Friedmann: the man who made the universe expand* (1993, Cambridge University Press).